\documentclass[12pt]{article}

\pdfoutput=1
\usepackage{graphicx}
\usepackage{amssymb}
\usepackage{amsmath}
\usepackage{color}
\usepackage{subfigure}
\usepackage[colorlinks=true,linkcolor=blue,citecolor=blue]{hyperref}

\begin{document}

\newcommand{\be}{\begin{equation}}
\newcommand{\ee}{\end{equation}}


\begin{titlepage}

\begin{flushright}
ICRR-Report-616-2012-5  \\
IPMU 12-0116
\end{flushright}

\vskip 2cm

\begin{center}

{\large \bf
Primordial black hole formation from an axion-like curvaton model
}

\vspace{1cm}

{Masahiro Kawasaki$^{(a, b)}$,  Naoya Kitajima$^{(a)}$ and Tsutomu T. Yanagida$^{(b)}$
}

\vskip 1.0cm

{\it
$^a$Institute for Cosmic Ray Research,
     University of Tokyo, Kashiwa, Chiba 277-8582, Japan\\
$^b$Kavli Institute for the Physics and Mathematics of the Universe, 
     University of Tokyo, Kashiwa, Chiba 277-8568, Japan\\
}

\vskip 1.0cm

\begin{abstract}
We argue that the existence of the cold dark matter is explained by primordial black holes.
We show that a significant number of primordial black holes can be formed in an axion-like curvaton model, in which the highly blue-tilted power spectrum of primordial curvature perturbations is achieved.
It is found that the produced black holes with masses $\sim 10^{20} -10^{38}~\mathrm{g}$ account for the present cold dark matter.
We also argue the possibility of forming the primordial black holes with mass $\sim 10^5 M_{\odot}$ as seeds of the supermassive black holes.
\end{abstract}

\end{center}

\end{titlepage}

\newpage



\vspace{1cm}


\section{Introduction} \label{intro}

The current cosmic microwave back ground (CMB) observations have revealed that the our present universe is filled with the unknown matter called dark matter, 
which cannot be explained within the framework of the well-established standard model of particle physics.
The observed density parameter for the cold dark matter (CDM)  is found by the WMAP \cite{Komatsu:2010fb} to be
\be
	\Omega_\mathrm{CDM} h^2 = 0.1126 \pm 0.0036,
	\label{DM_obs}
\ee
where $h$ is the dimensionless Hubble parameter defined via the present Hubble parameter: $H_0 = 100h~\mathrm{km~sec^{-1}~Mpc^{-1}}$.
In order to detect the dark matter, many experiments have been performed by now, but we have not found any meaningful signature yet.
Therefore it is one of the most important problems of modern cosmology and particle physics to answer what the dark matter is.

It is usually assumed that the dark matter is the weakly-interacting massive particles (WIMPs).
The supersymmetric (SUSY) model \cite{Martin:1997ns}, which is one of the most promising model beyond the standard model, naturally provides such WIMPs as the lightest supersymmetric particle (LSP).
Another promising candidate of dark matter is the axion, 
which is originally introduced to solve the strong CP problem in the standard model \cite{Peccei:1977hh}.
However, even if the SUSY particles or axion exist in the present universe, it may not be enough to explain the observed dark matter abundance depending on the model parameters.
In such a case, we are forced to demand another candidate for dark matter.
It is known that the primordial black holes (PBHs), the black holes formed in the early universe \cite{Zel'dovich&Novikov}, can behave like CDM.
In this paper, therefore, we argue the scenario in which the currently observed abundance of CDM is explained by PBHs.
 
PBHs are expected to be formed through the collapses of the high density regions caused by the large primordial density perturbations \cite{Carr:1975qj}.
Light PBHs with mass smaller than $10^{15}~\mathrm{g}$ are evaporated by now through the Hawking radiation \cite{Hawking:1974sw}, implying that only the PBHs with masses $M_\mathrm{BH} > 10^{15}~\mathrm{g}$ can survive and contribute to the CDM.
Furthermore, various cosmological and astrophysical constraints show that only the PBHs with mass $10^{17}~\mathrm{g} < M_\mathrm{BH} < 10^{27}~\mathrm{g}$ can be the dominant component of the current CDM \cite{Carr:2009jm}.
Although it is not easy to build the model in which a significant number of PBHs are formed, various models were proposed in the literature.
Focusing on the inflation models, for examples, PBH formation was proposed in double inflation models~\cite{Kawasaki:1997ju,Yokoyama:1998pt,Kawaguchi:2007fz,Frampton:2010sw} or running mass inflation models~\cite{Kohri:2007qn,Drees:2011hb}.

Another motivation to consider PBHs is the existence of supermassive black holes (SMBHs) at the center of galaxies~\cite{Kormendy:1995er,Magorrian:1997hw}.
The observation of quasars (QSO) reveals that the SMBHs with mass $M_\mathrm{BH} \approx 10^{9}~M_{\odot}$ exist at the redshift $z \approx 6$ \cite{Willott:2003xf}.
These black holes cannot be explained within the purely astrophysical mechanism, so we must rely on the primordial origin.
If PBHs with sufficiently large mass, $M_\mathrm{BH} \gtrsim 10^3 M_\odot$, can be formed in the early universe, they can play roles of the seeds of SMBHs~\cite{Duechting:2004dk,Kawasaki:2012kn}.

In order for PBHs to form through the primordial density perturbations, we need the strongly blue-tilted power spectrum of the curvature perturbations, 
which gives the large density perturbations at small scales while the large scale density perturbations are consistent with the CMB observation.
However, the observation indicates that the scale dependence of the power spectrum is slightly red-tilted at large scales.
This inconsistency is solved by employing a curvaton.
The curvaton was originally introduced to generate the primordial large scale curvature perturbations instead of the inflaton~\cite{Lyth:2001nq}.
In the curvaton model, a scalar field (called curvaton) acquires fluctuations during inflation and after inflation it decays into the standard model particles producing the adiabatic perturbation in the radiation dominated universe.

In this paper, we consider that the curvaton is responsible for generating only the small-scale curvature perturbations while the large-scale perturbations are generated by an inflaton.
After the decay of the curvaton, a significant number of PBHs can be formed through large density perturbations due to the curvaton.
A specific model for the PBH formation with curvaton was proposed in \cite{Yokoyama:1995ex}, where three scalar fields (including inflaton and curvaton) with ad hoc couplings among them evolve non-trivially during inflation and leads to large density perturbations at small scales.
Our mechanism for the PBH formation is completely different from that in Ref.~\cite{Yokoyama:1995ex}.
We consider an axion-like curvaton field whose nature is very crucial for the PBH formation.
Furthermore,  axion-like fields often appear in various particle physics theories.
We consider that one of such axion-like (curvaton) fields may play an important role for the PBH formation as studied in this paper. 
Ref.~\cite{Kohri:2007qn} also discussed PBH formation in curvaton model without concrete models.

The remainder of the paper is organized as follows.
In section \ref{model}, we show an axion-like curvaton model and we see the largely blue-tilted spectrum for the curvature perturbations is achieved.
In section \ref{PBH}, we consider the PBH formation within the axion-like curvaton model.
Section \ref{conc} is devoted to the conclusion.


\section{The axion-like curvaton model} \label{model}

\subsection{The potential of the curvaton} \label{potential}

In this section, we describe the axion-like curvaton model which was originally introduced in~\cite{Kasuya:2009up} (see also~\cite{Kasuya:1996ns}) 
as an axion model with extremely blue-tilted spectrum of the isocurvature perturbations.
The model is built in the framework of supersymmetry and has the following superpotential:
\be
	W = hS(\Phi \bar\Phi - f^2),
	\label{eq:super_pot}
\ee
where $\Phi$, $\bar\Phi$ and $S$  are chiral superfields whose $R$-charges are $+1$, $-1$ and $+2$ respectively, $f$ is some energy scale and $h$ is a dimensionless coupling constant.
Here we assume that the model has a global $U(1)$ symmetry and $\Phi$, $\bar\Phi$ and $S$ have charges $+1$, $-1$ and $0$, respectively.
In the limit of the global SUSY, the scalar potential is derived from (\ref{eq:super_pot}) as 
\be
	V = h^2 | \Phi \bar\Phi - f^2|^2 + h^2 |S|^2 ( |\Phi|^2 + |\bar\Phi|^2 ),
	\label{scalar_pot}
\ee
where, the scalar components are denoted by the same symbols as the superfields. 
Provided that $|S| < f$ is satisfied, $S$ tends to the origin and $\Phi$ and $\bar\Phi$ are settled on the flat direction satisfying
\be
	\Phi \bar\Phi = f^2 ~~~\text{with}~~~S = 0, 
	\label{flat_direction}
\ee
which makes the scalar potential (\ref{scalar_pot}) vanish.
Hereafter, we assume that the flat condition (\ref{flat_direction}) is always satisfied.
Including the supergravity effects, the Hubble-induced mass terms are added to the scalar potential \cite{Dine:1995kz} as
\be
	V_H = c_1 H^2 |\Phi|^2 + c_2 H^2 |\bar\Phi|^2 + c_S H^2 |S|^2,
	\label{Hubble_mass}
\ee
where $c_1$, $c_2$ and $c_S$ are numerical constants assumed to be real, positive and of order unity.
In addition, there also exist the low energy SUSY breaking terms, 
\be
	V_m = m_1^2 |\Phi|^2 + m_2^2 |\bar\Phi|^2 + m_S^2 |S|^2, 
\ee
where $m_1$, $m_2$ and $m_S$ are soft masses of order of the gravitino mass.
Here, because we are interested in the inflationary epoch, we neglect the low energy SUSY breaking mass terms.
Thus, the flat direction is lifted by only the Hubble-induced mass terms (\ref{Hubble_mass}) and the minimums of $\Phi$ and $\bar\Phi$ are determined as
\be
	|\Phi|_\mathrm{min} \simeq \bigg( \frac{c_2}{c_1} \bigg)^{1/4} f,~~~
	|\bar\Phi|_\mathrm{min} \simeq \bigg( \frac{c_1}{c_2} \bigg)^{1/4} f.
	\label{min}
\ee

Now, we decompose the complex scalar fields into the radial and the phase components as 
\be
	\Phi = \frac{1}{\sqrt 2} \varphi \exp( i\theta_+ ),~~~ \bar\Phi = \frac{1}{\sqrt 2} \bar\varphi \exp (i\theta_- ).
\ee
Then, along the flat direction, the massless direction is found as a linear combination of the phases, $\theta = (\theta_+ - \theta_-) / 2$.
Without loss of generality, we can take $\varphi \gg \bar\varphi$ as the initial condition and neglect the dynamics of $\bar\varphi$ in the early epoch~\cite{Kasuya:1996ns}, 
so we follow the dynamics of only the complex scalar field $\Phi = \varphi e^{i\theta} / \sqrt 2$
whose potential is given by
\be
	V_\varphi = \frac{1}{2} c H^2 \varphi^2.
	\label{varphi_pot}
\ee
Note that since the mass is comparable to the Hubble parameter, $\varphi$ rolls down the potential somewhat rapidly during inflation.

In our model, the curvaton is defined as the phase component of $\Phi$.
Note that the curvaton is well-defined only after $\varphi$ reaches the minimum ($\varphi_{\rm min}$) and denoted as $\sigma = \varphi_{\rm min} \theta \sim f \theta$.
Here, we assume that the $U(1)$ symmetry is broken by some non-perturbative effect and $\sigma$ has the following potential in the low energy universe like the axion:
\be
	V_\sigma = \Lambda^4 \bigg[ 1 - \cos \bigg( \frac{\sigma}{f} \bigg) \bigg] \simeq \frac{1}{2} m_\sigma^2 \sigma^2, 
	\label{curv_pot}
\ee
where the second equality holds near the minimum $\sigma_\mathrm{min} = 0$ and the curvaton mass is defined as $m_\sigma  = \Lambda^2 / f$.
After the Hubble parameter becomes smaller than the curvaton mass, the curvaton field starts to oscillate coherently with the initial amplitude $\sigma_i \sim f \theta$ 
and behaves as matter. 

Let us derive the ratio $r$ of the curvaton energy density to that of the radiation at the time of the curvaton decay.
In order to estimate this, we consider the two cases : (i) the reheating occurs after the curvaton starts to oscillate and (ii) the curvaton starts to oscillate after the reheating.
In case (i), which is denoted as $m_\sigma > \Gamma_I$ ($\Gamma_I$ : decay rate of the inflaton), we estimate $r$ as
\be
	r = \frac{\rho_\sigma(t_\mathrm{dec})}{\rho_r (t_\mathrm{dec})} = \frac{\rho_\sigma(t_R)}{\rho_r (t_R)} \frac{T_R}{T_\mathrm{dec}} 
	\simeq \frac{1}{6} \bigg( \frac{f \theta}{M_P} \bigg)^2 \frac{T_R}{T_\mathrm{dec}} ~~~\text{for}~~m_\sigma \gtrsim \frac{T_R^2}{M_P}, 
	\label{reheat1}
\ee
and, on the other hand, in case (ii), $m_\sigma < \Gamma_I$, we obtain
\be
	r  = \frac{\rho_\sigma(t_\mathrm{osc})}{\rho_r (t_\mathrm{osc})} \frac{T_\mathrm{osc}}{T_\mathrm{dec}}
	\simeq \frac{1}{6} \bigg( \frac{f \theta}{M_P} \bigg)^2 \frac{T_\mathrm{osc}}{T_\mathrm{dec}}~~~\text{for}~~m_\sigma \lesssim \frac{T_R^2}{M_P},
	\label{reheat2}
\ee
where the subscripts dec,  osc and $R$ correspond to the epochs of the curvaton decay, the curvaton oscillation and the reheating, respectively. 
Hereafter, we consider only the case of $r \leq 1$.

The curvaton decays when the Hubble parameter becomes equal to the decay rate of the curvaton and the decay temperature of the curvaton is determined from the decay rate.
Here we assume that the interaction of the curvaton with its decay product is suppressed by $f$ like an axion, so we denote the decay rate of the curvaton as
\be
	\Gamma_\sigma = \frac{\kappa^2}{16 \pi} \frac{m_\sigma^3}{f^2},
	\label{decay_rate_curv}
\ee
where $\kappa$ is a dimensionless numerical constant assumed to be real, positive and smaller than 1.
Then, the decay temperature of the curvaton is given by 
\be
	T_\mathrm{dec} = 0.5 \bigg( \frac{g_*}{100} \bigg)^{-1/4} ( \Gamma_\sigma M_P)^{1/2},
\ee
where $g_*$ is the relativistic degrees of freedom.

\subsection{Generating the curvature perturbation} \label{curv}

To create a significant number of PBHs from the primordial density perturbations, we require the extremely blue spectrum with spectral index $n_s \gtrsim 2$ as we will see later.
However, such a large spectral index is already ruled out by the CMB observation.
In order not to contradict the observation, then, we build the model in which the almost scale-invariant large-scale curvature perturbations are generated by an inflaton and 
the small-scale curvature perturbations which are free from the CMB constraint are generated by the curvaton.
Here, we investigate the possibility of PBH formation in the axion-like curvaton model introduced above. 

The power spectrum of curvature perturbations is the sum of the contributions from the inflaton and the curvaton, which is written as 
\be
	\mathcal P_\zeta (k) = \mathcal P_{\zeta, \mathrm{inf}} (k) + \mathcal P_{\zeta, \mathrm{curv}} (k).
	\label{P_zeta_sum}
\ee
As mentioned above, the power spectrum is dominated by the first term in rhs. of (\ref{P_zeta_sum}) for small $k$ and by the second term for large $k$.
It is quite reasonable that the contribution from the inflaton is dominant until the perturbation scale at least $k \sim 1~\mathrm{Mpc}^{-1}$ leaves the horizon.
For later convenience, we define $k_\mathrm{c}$ as
\be
	\mathcal P_{\zeta,\mathrm{curv}} (k_\mathrm{c}) 
	= \mathcal P_{\zeta,\mathrm{inf}} (k_\mathrm{c}) 
	\simeq 2 \times 10^{-9} ~~~
	\label{kc_def}
\ee
where we have used the CMB normalization \cite{Komatsu:2010fb} for the spectrum of the large-scale curvature perturbations.
Using this definition, our requirement is  denoted as 
\be
	\mathcal P_{\zeta, \mathrm{curv}} (k) < \mathcal P_{\zeta, \mathrm{inf}} (k) 
	\sim 2 \times 10^{-9} ~~~ 
	\text{for}~~~ k ~\lesssim ~ k_\mathrm{c}~~{\rm and} ~~
	k_c ~\gtrsim ~1~\mathrm{Mpc}^{-1}.
	\label{P_zeta_condition}
\ee

Let us consider the power spectrum from the fluctuation of the curvaton field.
From the definition of the curvaton, $\sigma = \varphi_{\rm min}\theta$, the density perturbation of the curvaton is given by
\be
	\frac{\delta\rho_\sigma}{\rho_{\sigma,0}} \simeq \frac{2\delta\sigma}{\sigma_0} = \frac{2\delta\theta}{\theta_0},
\ee
where we decompose each field into the homogeneous part  and the small perturbation: $X = X_0(t) + \delta X(t,\vec{x})$.
Focusing on the super-horizon Fourier mode of $\delta\theta$, $\delta\theta / \theta_0$ is conserved 
because the masses of both $\theta_0$ and $\delta\theta$ are much smaller than the expansion rate of the universe \cite{Lyth:1991ub}.
This means that the resultant power spectrum remembers the fluctuation of $\theta$ at the horizon exit, written as $\mathcal{P}^{1/2}_{\delta\theta}(k) \simeq H_\mathrm{inf}/(2\pi\varphi_0(k))$, 
where the argument $k$ entering in $\varphi_0$ denotes the value when the scale $k$ leaves the horizon.
Thus, the power spectrum of the density perturbation for the curvaton is expressed as 
\be
	\mathcal{P}^{1/2}_{\delta,\mathrm{curv}}(k) = \frac{2 \mathcal{P}_{\delta\theta}^{1/2}(k)}{\theta} \simeq \frac{H_\mathrm{inf}}{\pi \varphi(k) \theta}.
\ee
Here and hereafter, we drop the subscript 0 to express the homogeneous value.
From the above, the spectrum of the curvature perturbation from the curvaton is calculated as \cite{Lyth:2001nq}
\be
	\mathcal{P}_{\zeta,\mathrm{curv}} (k) = \bigg( \frac{r}{4+3r} \bigg)^2 \mathcal{P}_{\delta,{\rm curv}}
	= \bigg( \frac{2r}{4+3r} \bigg)^2 \bigg( \frac{H_\mathrm{inf}}{2 \pi \varphi (k) \theta} \bigg)^2
	\label{P_zeta_curv}
\ee
Note that, after $\varphi$ reaches the minimum, the power spectrum takes the constant value given by
\be
	\mathcal P_{\zeta,\mathrm{curv}} (k) = \mathcal P_{\zeta, \mathrm{curv}} (k_*) 
	\approx \bigg( \frac{2r}{4 + 3r} \bigg)^2 \bigg( \frac{ H_\mathrm{inf}}{2 \pi f \theta} \bigg)^2~~~\text{for}~~~ k > k_*,
	\label{P_zeta_curv2}
\ee
where $k_*$ is defined as the scale leaving the horizon at the time $\varphi$ reaches the minimum $\sim f$.
With use of the curvaton spectral index $n_\sigma$, the scale dependence of the power spectrum of the curvature perturbation is expressed as 
\be
	\mathcal P_{\zeta, \mathrm{curv}} (k) = \mathcal P_{\zeta, \mathrm{curv}} (k_\mathrm{c}) \bigg( \frac{k}{k_\mathrm{c}} \bigg)^{n_\sigma - 1} ~~~\text{for}~~~ k \leq k_*.
	\label{P_zeta_curv3}
\ee
Combining this and the $\varphi$ dependence,
\be
	\mathcal P_{\zeta, \mathrm{curv}} (k) 
	= \mathcal P_{\zeta, \mathrm{curv}} (k_\mathrm{c}) 
	\bigg( \frac{\varphi (k_\mathrm{c})}{\varphi (k)} \bigg)^2,
	\label{varphi-dependence_of_P_zeta}
\ee
we obtain the relation 
\be
	k = k_\mathrm{c} 
	\bigg( \frac{\varphi (k_\mathrm{c})}{\varphi (k)} \bigg)^{2/(n_\sigma - 1)} 
	~~~\text{for}~~~ k \leq k_*.
	\label{k-varphi}
\ee

The spectral index of the curvaton is calculated by solving the equation of motion of $\varphi$ with potential (\ref{varphi_pot}), 
\be 
    \ddot\varphi +3H\dot\varphi + cH^2\varphi = 0,
\ee
whose solution during inflation ($H \simeq $ const.) is given by 
\be
    \varphi ~\propto~ e^{-\lambda Ht} ~\propto~ k^{-\lambda}~~~{\rm with}~~ 
    \lambda = \frac{3}{2}-\frac{3}{2}\sqrt{1-\frac{4}{9}c}.
\ee
Together with (\ref{P_zeta_curv3}) and (\ref{varphi-dependence_of_P_zeta}) the spectral index is given by
\be
	n_\sigma -1 = 3-3 \sqrt{1-\frac{4}{9}c},
\ee
so we can obtain the extremely blue spectrum such as $n_\sigma \sim 2$~--~4 with appropriate choice of $c$ \cite{Kasuya:2009up}.


\section{The PBH formation} \label{PBH}

In this section, we consider the formation of PBHs in our model.
It is well-known that PBHs can be formed by collapse of overdensity regions in the radiation-dominated universe 
and their mass is as large as the horizon mass at the formation time \cite{Niemeyer:1997mt,Green:1999xm,Shibata:1999zs}, which is given by
\be
	\begin{split}
	M_\mathrm{BH} = \frac{4\pi}{3} \rho_r H^{-3} &\simeq 0.05 M_\odot \bigg( \frac{g_*}{100} \bigg)^{-1/2} \bigg( \frac{T_f}{\mathrm{GeV}} \bigg)^{-2} \\[1mm]
	&\simeq 1 \times 10^{13} M_\odot \bigg( \frac{g_*}{100} \bigg)^{-1/6} \bigg( \frac{k_f}{\mathrm{Mpc}^{-1}} \bigg)^{-2} ,
	\end{split}
	\label{M_BH}
\ee
where $\rho_r$ represents the energy density of the radiation and the subscript  $f$ represents the time of the PBH formation.
Here we assume $r \leq 1$ which means that PBHs are formed in radiation dominated universe after the curvaton decay (see also footnote~\ref{foot:before_decay}).
PBHs with $M_\mathrm{BH} > 10^{15}~\mathrm{g}$ do not evaporate through the Hawking radiation \cite{Hawking:1974sw} until now 
and their abundance can contribute to the present CDM density.
The current density parameter for such PBHs is calculated as 
\be
	\Omega_\mathrm{PBH} h^2 = \frac{\rho_\mathrm{PBH,eq}}{\rho_\mathrm{tot,eq}} \Omega_{\rm m} h^2 \simeq 5 \times 10^7 \beta \bigg( \frac{M_\odot}{M_\mathrm{BH}} \bigg)^{1/2},
\ee
where the subscript eq corresponds the time of the matter-radiation equality and $\Omega_{\rm m} \simeq 0.13h^{-2}$ is the density parameter for matter today. 
$\beta$ is defined as the energy density fraction of the PBHs at the PBH formation, which is denoted as $\beta \equiv \rho_\mathrm{PBH}(t_f) / \rho_\mathrm{tot} (t_f)$.
Various cosmological and astrophysical constraints are imposed on $\beta$~\cite{Carr:2009jm}, from which 
only the PBHs with mass $\sim 10^{17}~\mathrm{g}$ - $10^{27}~\mathrm{g}$ can be the dominant component of the dark matter.
For such PBHs, the constraint on $\beta$ comes from the current observational value of the dark matter density (\ref{DM_obs}), which implies
\be
	\beta < 3 \times 10^{-11} \bigg( \frac{M_\mathrm{BH}}{M_\odot} \bigg)^{1/2}.
	\label{constraint_on_beta}
\ee

\begin{figure}[t]
\centering
\includegraphics [width = 8cm, clip]{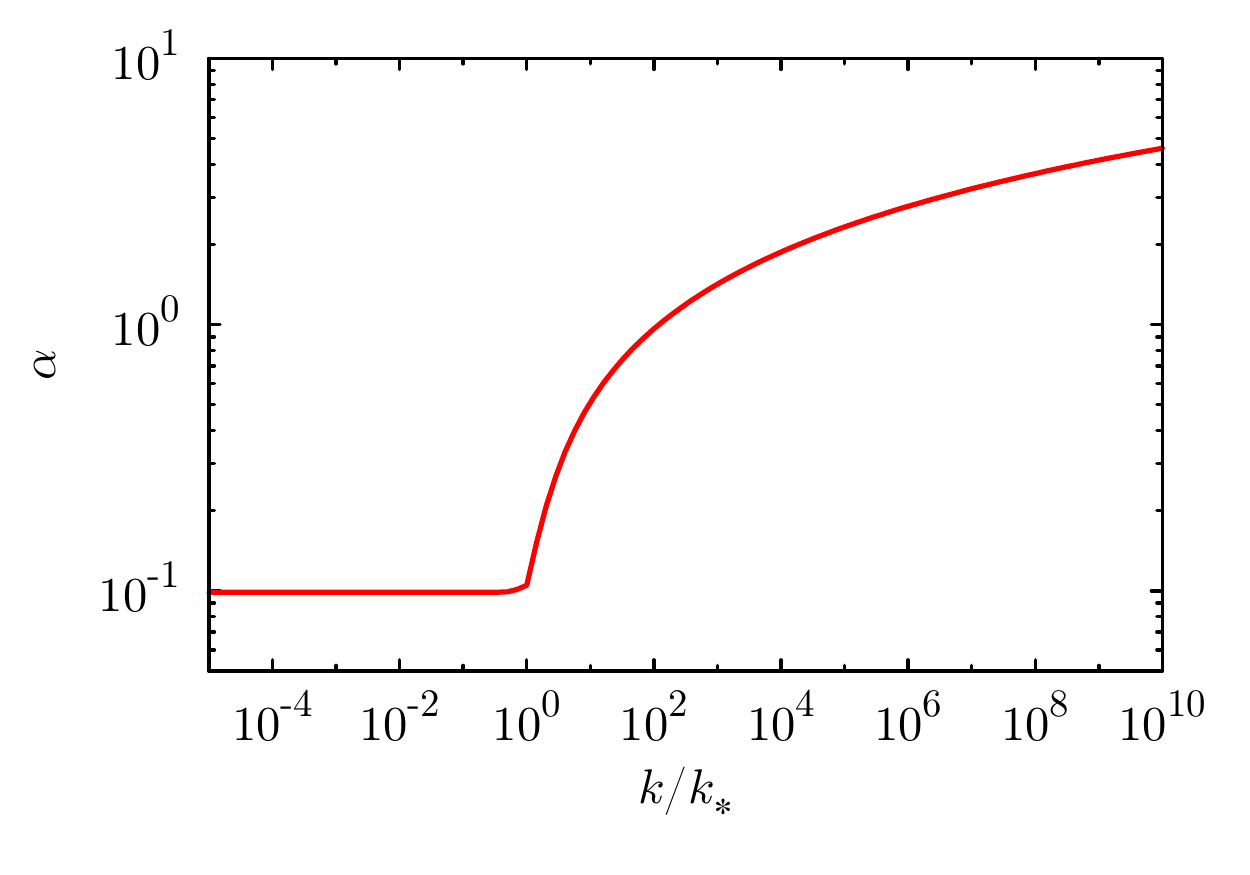}
\caption{
	The ratio of the smoothed variance of density perturbation to the power spectrum of curvature perturbation are shown.
	The horizontal axis is the wave number corresponding to the smoothing scale as $k=R^{-1}$ divided by $k_*$.
	This curve is independent of $n_\sigma$.
}
\label{Fig1}
\end{figure}

Assuming that the PBHs are created by the collapse of overdensity regions with primordial gaussian density perturbations, $\beta$ is estimated as \cite{Green:1997sz}
\be
	\beta \approx \sqrt{\langle \delta^2 \rangle} \exp \bigg( -\frac{1}{18 \langle \delta^2 \rangle} \bigg),
	\label{beta}
\ee
where $\delta = \delta \rho / \rho$ is the density contrast and $\langle \delta^2 \rangle$ is its variance.
The variance of the density perturbations is related to the power spectrum of curvature perturbations, 
and, in the comoving gauge in which the curvature perturbation is expressed as $\mathcal R$, which coincides with $\zeta$ well outside the horizon, 
the following relation is known:
\be
	\mathcal P_\delta (k) = \frac{4(1+w)^2}{(5+3w)^2} \mathcal P_{\mathcal R} (k), 
\ee
at the time the scale $k$ leaves the horizon \cite{Liddle&Lyth}.
$w$ is determined from the relation between the pressure and the energy density of the cosmic fluid $P = w \rho$ and 
it takes 1/3 in radiation dominated universe.
The variance of the density perturbation smoothed over the scale $R$ is estimated as
\be
	\langle \delta^2 (R) \rangle = \int^{\infty}_0 W^2(kR) \mathcal P_\delta (k) \frac{dk}{k},
\ee
where $W(kR)$ is the window function in Fourier space.
Assuming the Gaussian window function $W(kR) = \exp(-k^2 R^2/2)$ and taking into account $\mathcal P_\zeta \approx \mathcal P_{\mathcal R}$ and (\ref{P_zeta_curv2}) and (\ref{P_zeta_curv3}),
we can approximate the variance as
\be
	\langle \delta^2 (R) \rangle = \frac{8}{81} \mathcal P_{\zeta,\mathrm{curv}}(k_*) \big[ (k_*R)^{-(n_\sigma -1)} \gamma ( (n_\sigma-1)/2 , k_*^2 R^2 ) 
	+ E_1( k_*^2 R^2 ) \big],
	\label{smooth_var}
\ee
where $\gamma(a,x)$ and $E_1$ are defined as
\be
	\gamma(a,x) = \int^x_0 t^{a-1}e^{-t}dt,~~~{\rm and} ~~~~
	E_1(x) = \int^\infty_x \frac{e^{-t}}{t}dt.
\ee
If we wrire
\be
	\langle \delta^2 (k^{-1}) \rangle = \alpha \mathcal P_{\zeta, \mathrm{curv}} (k),
	\label{var}
\ee
the numerical coefficient $\alpha$ is taken to be $0.1 \mathchar`- 4$ as shown in Fig.~\ref{Fig1}.

\begin{figure}[t]
\centering
\subfigure[]{
	\includegraphics [width = 7.5cm, clip]{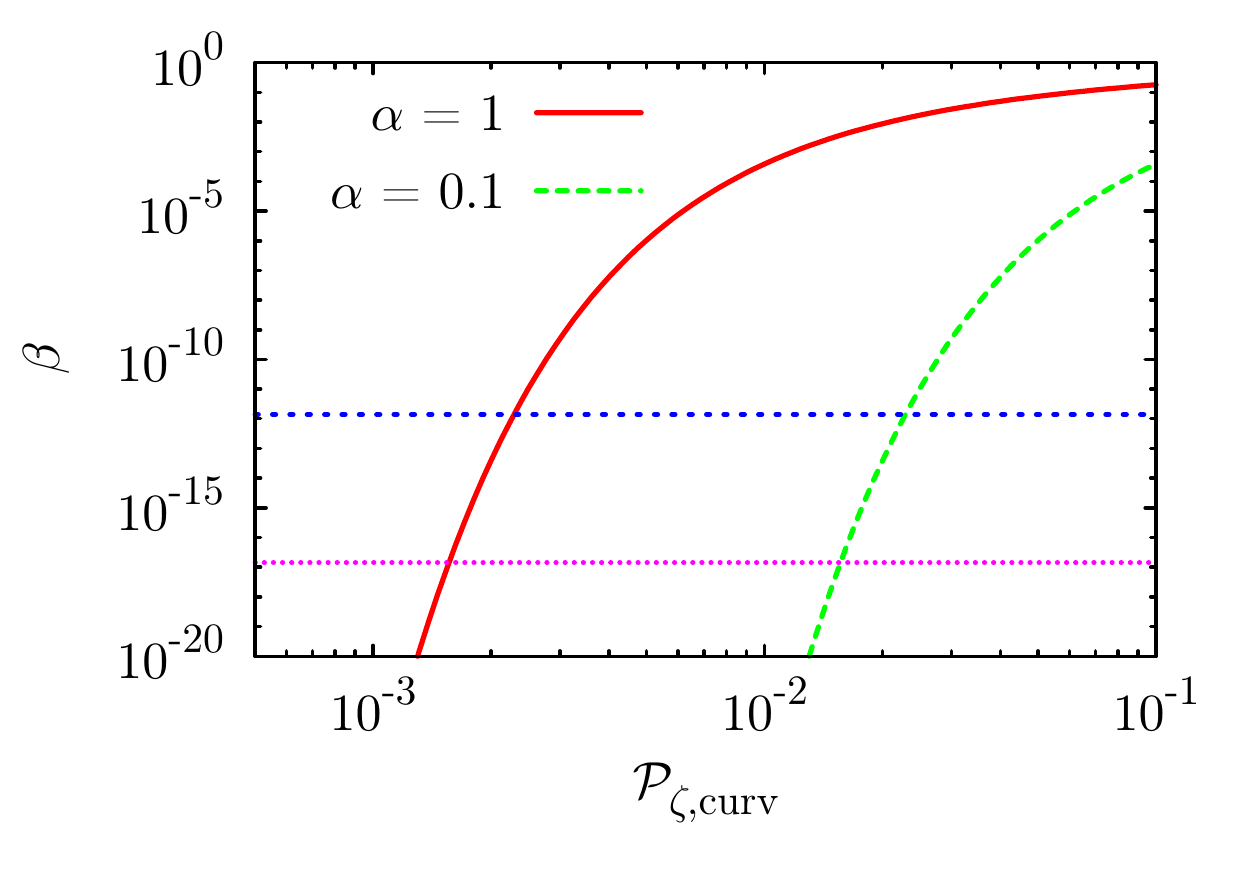}
	\label{Fig2a}
}
\subfigure[]{
	\includegraphics [width = 7.5cm, clip]{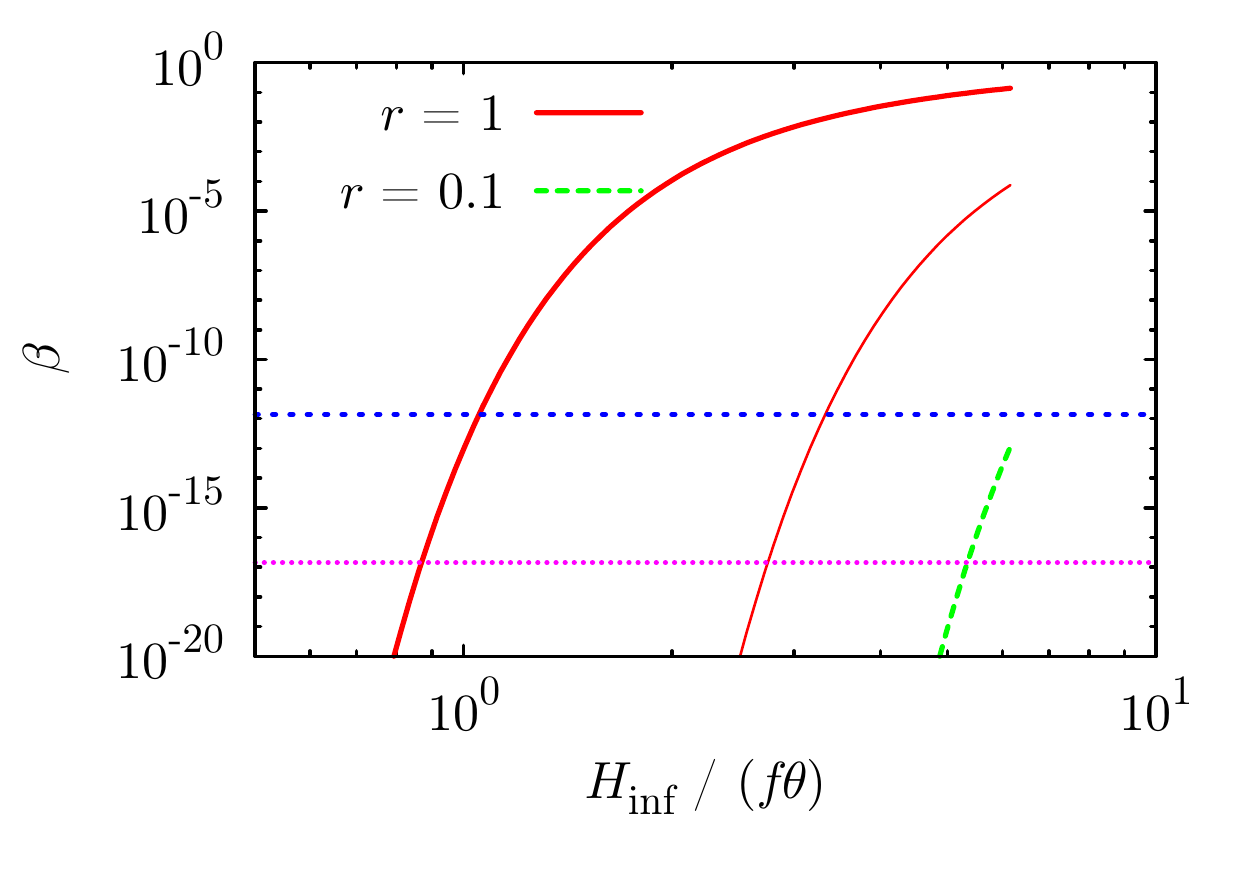}
	\label{Fig2b}
}
\caption{
The energy density fraction of the PBH at the formation is shown.
The horizontal axis correspond to $\mathcal P_{\zeta,\mathrm{curv}}$ in Fig.~\ref{Fig2a} and $H_\mathrm{inf} / f \theta$ in Fig.~\ref{Fig2b}.
In Fig.~\ref{Fig2a}, the solid red line and the dashed green line correspond to $\alpha = 1$ and $\alpha = 0.1$ respectively.
In Fig.~\ref{Fig2b}, the thick (thin) solid red line corresponds to $r = 1$ and $\alpha=1~(0.1)$ 
and the thick dashed green line corresponds to $r=0.1$ and $\alpha = 1$.
Breaking point of each line in Fig.~\ref{Fig2b} corresponds to $\delta \sigma / \sigma = 1$.
The dotted blue line (the small-dotted magenta line) corresponds to the upper limit in the case of $M_\mathrm{BH} = 10^{27}~(10^{17})~\mathrm{g}$, 
which comes from the current observational value of the CDM density parameter : $\Omega_\mathrm{CDM} = 0.23$.
}
\label{Fig2}
\end{figure}

We show the energy density fraction of PBH in terms of $\mathcal P_{\zeta,\mathrm{curv}}$ in Fig.~\ref{Fig2a} in the case of $\alpha = 1$ (solid red line) and $\alpha = 0.1$ (dashed green line).
The dotted blue line (the small-dotted magenta line) corresponds to the upper limit in the case of $M_\mathrm{BH} = 10^{27}~(10^{17})~\mathrm{g}$, 
which comes from the current observation of the CDM density.
In order for PBHs to be the dominant component of dark matter, 
the required value of curvature perturbation is $\mathcal P_{\zeta,\mathrm{curv}} \sim 2 \times 10^{-3}~(2 \times 10^{-2})$ for $\alpha = 1~(0.1)$.
Substituting (\ref{P_zeta_curv}) and (\ref{var}) into (\ref{beta}) and taking $\varphi(k) \sim f$,  
the constraint (\ref{constraint_on_beta}) is rewritten in terms of $H_{\mathrm{inf}}/(f \theta)$ shown in Fig.~\ref{Fig2b}.
In this figure, the thick (thin) solid red line corresponds to $r = 1$ and $\alpha=1~(0.1)$ and the thick dashed green line corresponds to $r=0.1$ and $\alpha = 1$.
The breaking point of each line corresponds to the point at whch the quantum fluctuation of the curvaton, $\delta \sigma = H_\mathrm{inf}/2 \pi$, becomes $f$.
If $H_\mathrm{inf}/2 \pi > f$, the amplitude of the quantum fluctuations of $S$ overtakes the critical value $f$, which invalidates our underlying assumption (\ref{flat_direction}).
From Fig.~\ref{Fig2b} we need $ r \sim 1$ and  $f \theta \sim H_\mathrm{inf}$ to account for the present dark matter abundance.\footnote{
	If we allow $r>1$, PBHs can be formed after the curvaton starts to dominate the universe.
	The PBH formation in matter dominated universe is discussed in \cite{Khlopov:1980mg} and the initial energy fraction of PBH is estimated as 
	$\beta \simeq 2 \times 10^{-2} \langle \delta^2 \rangle^{13/4}$.
	We have found $\mathcal{P}_{\zeta} \sim 2 \times 10^{-4}$ and $f \theta \sim 10 H_{\rm inf}$ to explain the present dark matter abundance.
	We also note that there is a non-negligible effect from the non-Gaussianity in the case of $r \gg1$.
	In such a case, the non-Gaussianity parameter $f_\mathrm{NL}$ becomes negative and the resultant PBH abundance becomes too small to be the dominant component of the CDM \cite{Byrnes:2012yx}.
}

\begin{figure}[t]
\centering
\includegraphics [width = 10cm, clip]{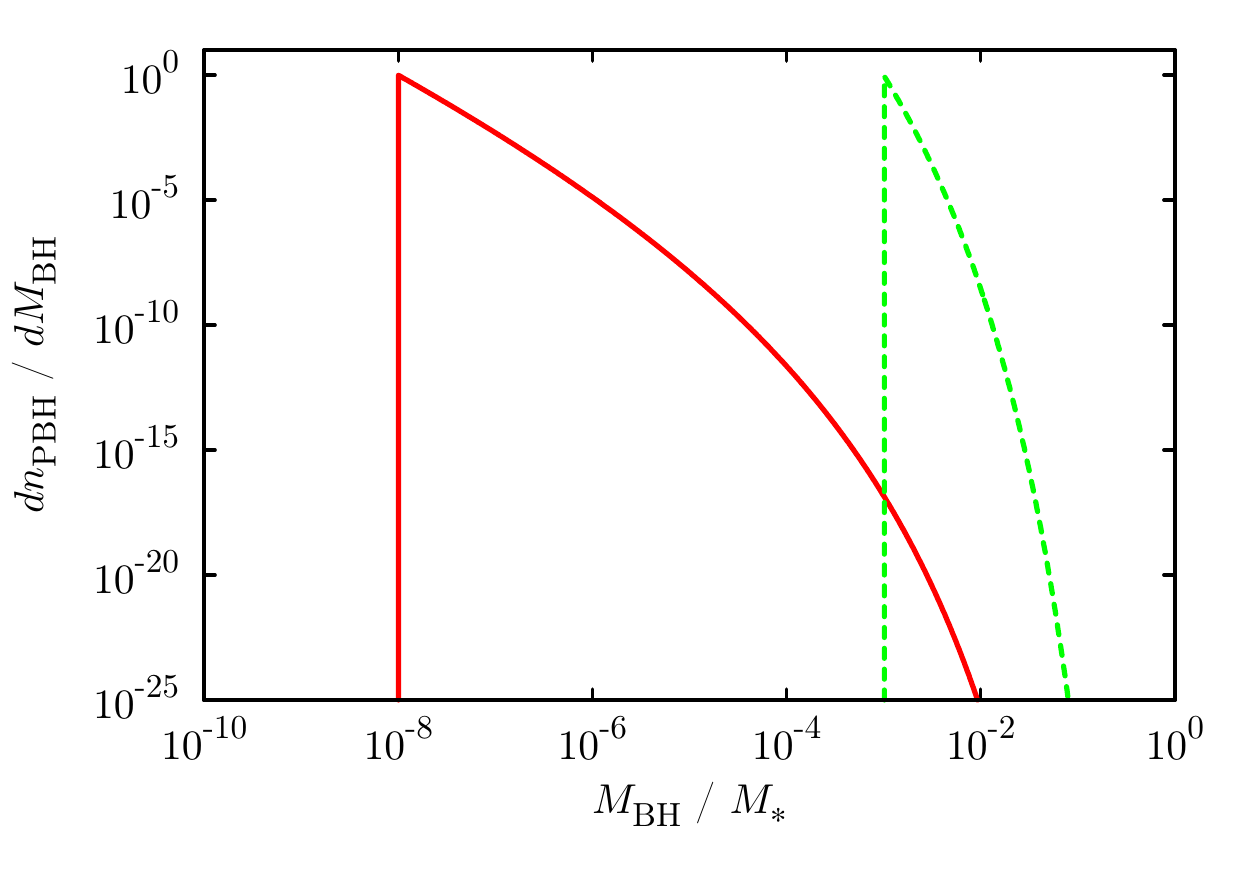}
\caption{
	The mass spectrum of PBH, $dn_\mathrm{PBH}/dM_\mathrm{BH}$, is shown.
	The solid red line and dashed green line corresponds to $M_\mathrm{min}/M_* = 10^{-8}$ and $M_\mathrm{min}/M_* = 10^{-3}$ respectively and they are normalized by the their own peak values.
	They are independent of $n_\sigma$.
}
\label{Fig3}
\end{figure}

Now let us estimate the mass spectrum of PBHs in the present model. 
This is especially impotent for SMBHs since, taking into account the merging and accretion events prior to the formation of SMBHs, 
the mass spectrum of primordial seeds of SMBHs is required to have a sharply peaked shape \cite{Bean:2002kx}.
With $R=k^{-1}$ and Eq.(\ref{M_BH}) we rewrite the smoothed variance (\ref{smooth_var}) in terms of PBH masses as
\be
	\langle \delta^2(M_\mathrm{BH}) \rangle = \frac{8}{81} \mathcal P_{\zeta,\mathrm{curv}} (k_*) 
	\bigg[ \bigg( \frac{M_*}{M_\mathrm{BH}} \bigg)^{(n_\sigma -1)/2} \gamma \bigg( \frac{n_\sigma -1}{2}, \frac{M_\mathrm{BH}}{M_*} \bigg) + E_1 \bigg( \frac{M_\mathrm{BH}}{M_*} \bigg) \bigg],
	\label{mass_variance}
\ee
where $M_*$ is the mass of PBH formed when the scale $k_*$ enters the horizon.
Using (\ref{mass_variance}), we can calculate the mass function, which is defined as the number of PBHs per comoving volume 
whose mass range is $M_\mathrm{BH}$ $\sim$ $M_\mathrm{BH}+dM_\mathrm{BH}$, as~\cite{Kim:1996hr} : 
\be
	\frac{dn_\mathrm{PBH}}{dM_\mathrm{BH}} = \sqrt{\frac{1}{18\pi}} \frac{\bar\rho_r (t_*)}{M_\mathrm{BH}^2} \bigg( \frac{M_*}{M_\mathrm{BH}} \bigg)^{1/2} 
	\bigg| \frac{d\ln\langle \delta^2(M_\mathrm{BH})\rangle}{d\ln M_\mathrm{BH}} \bigg| \frac{\beta (M_\mathrm{BH})}{\langle \delta^2(M_\mathrm{BH})\rangle},
\ee
where $\bar\rho_r(t_*)$ is the radiation energy per comoving volume when the scale $k_*$ enters the horizon and $\beta (M_\mathrm{BH})$ is the density fraction of PBH whose mass is $M_\mathrm{BH}$.
Since we assume that PBHs are formed after the curvaton decays, the mass spectrum has the lower cutoff $M_\mathrm{min}$, which corresponds to
the mass of PBH formed just after the curvaton decays.\footnote{
	PBH formation before the curvaton decays in radiation dominated universe may also be possible.
	In such a case, however, $\mathcal{P}_\delta$ is suppressed through the factor $(\rho_\sigma/\rho_r)^2$ at the formation.
	Since the number of produced PBHs is very sensitive to $\mathcal{P}_\delta$ and exponentially suppressed for small $\mathcal{P}_\delta$, 
	the number of those PBHs produced before the curvaton decay may be negligibly small.
	Thus, even if we include the above effect, the mass spectrum (Fig.~\ref{Fig3}) may slightly spread around the cutoff and nothing is affected in our discussion.
\label{foot:before_decay}}
The mass spectrum of PBHs is shown in Fig.~\ref{Fig3}.
The solid red and dashed green lines correspond to $M_\mathrm{min}/M_* = 10^{-8}$ and $M_\mathrm{min}/M_* = 10^{-3}$ respectively and they are normalized by the their own peak values.
The mass spectrum depends only on $M_\mathrm{min}$ and it is independent of $n_\sigma$.
It is clear that  the dominant contribution to the energy density of PBHs comes from the smaller mass PBHs, 
so the constraint on the initial PBH abundance should be applied to the PBHs with $M_\mathrm{min}$.
In particular, for $M_\mathrm{min}/M_* = 10^{-8}$, it is seen that the number of PBHs with mass larger than $\sim 10^{-4}M_*$ decreases drastically.
This is due to the sudden decreasing of $\alpha$, which implies the sudden decreasing of $\langle \delta^2 \rangle$, for $k \lesssim 10^2 k_*$ (see Fig.~\ref{Fig1}).
Thus we can obtain a very narrow mass spectrum by tuning $M_\mathrm{min}/M_*$ as $10^{-3}$--$10^{-2}$, which is required to explain SMBHs.

Now, we investigate the parameters allowing the formation of PBHs which eventually becomes the dominant component of the CDM.
We also impose several conditions to build the viable scenario, which are listed below.
\begin{itemize}
\item
Going back to the time when the pivot scale $k_p = 0.002~\mathrm{Mpc^{-1}}$ leaves the horizon, $\varphi$ should be smaller than the Planck scale.
Using (\ref{k-varphi}) and  taking into account $f \approx H_\mathrm{inf} / \theta$ and $\varphi (k_\mathrm{c}) \approx 10^{3}f$, we get
\be
	H_\mathrm{inf} < 2^{(n_\sigma - 1)/2} 10^{-3 (n_\sigma + 1)/2} \theta M_P \bigg( \frac{k_\mathrm{c}}{\mathrm{Mpc^{-1}}} \bigg)^{-(n_\sigma-1)/2}.
	\label{constraint_on_H_inf}
\ee
Combining this with the constraint from the tensor-to-scalar ratio \cite{Komatsu:2010fb}, $H_\mathrm{inf} < 5 \times 10^{-5} M_P$, we get 
\be
	H_\mathrm{inf} < \min \big[~ 2^{(n_\sigma - 1)/2} 10^{-3 (n_\sigma + 1)/2} \theta M_P,~5 \times 10^{-5}M_P ~\big],
	\label{constraint_on_H_inf2}
\ee
where we set $k_\mathrm{c} = 1~\mathrm{Mpc}^{-1}$.
\item
The mass of PBHs formed when the scale $k_*$ reenter the horizon, $M_*$,  is calculated with use of (\ref{kc_def}),(\ref{P_zeta_curv3}) and (\ref{M_BH}). 
Since $M_*$ is larger than the minimum mass of PBHs, we obtain the following condition:  
\be
	M_\mathrm{min} < M_{*}= 2 \times 10^{46 - 12/(n_\sigma - 1)} ~ \mathrm{g} \bigg( \frac{g_*}{100} \bigg)^{-1/6}\bigg( \frac{k_\mathrm{c}}{\mathrm{Mpc^{-1}}} \bigg)^{-2} 
	\bigg( \frac{\mathcal P_{\zeta,\mathrm{curv}}(k_*)}{2 \times 10^{-3}} \bigg)^{-2/(n_\sigma -1)}
	\label{M_BH_condition1}
\ee
\item
The curvaton should decay before the Big Bang nucleosynthesis (BBN), that is $T_\mathrm{dec} > 1~\mathrm{MeV}$.
From (\ref{M_BH}), the minimum mass of PBHs is related to $T_\mathrm{dec}$ as
\be
	T_\mathrm{dec} \simeq 1 \times 10^3 ~\mathrm{GeV} \bigg( \frac{g_*}{100} \bigg)^{-1/4} \bigg( \frac{10^{26}~\mathrm{g}}{M_\mathrm{min}} \bigg)^{1/2}.
	\label{Tdec-Mmin}
\ee
Hence the minimum mass of PBHs is constrained as
\be
	M_\mathrm{min} \lesssim  1 \times 10^{38}~\mathrm{g} \bigg( \frac{g_*}{100} \bigg)^{-1/2},
	\label{M_BH_condition2}
\ee
which is combined with (\ref{M_BH_condition1}) and we get 
\be
	M_\mathrm{min} \lesssim \min \big[ ~2 \times 10^{46 - 12/(n_\sigma - 1)} ~ \mathrm{g},~1 \times 10^{38}~\mathrm{g} ~\big], 
	\label{M_BH_cons}
\ee
where we set $g_* \approx 100$, $k_c=1~\mathrm{Mpc}^{-1}$ and $\mathcal P_{\zeta,\mathrm{curv}}(k_*) = 2 \times 10^{-3}$.
\item
The reheating temperature is also constrained.
In the case of $\Gamma_I < m_\sigma$, from (\ref{reheat1}) and (\ref{Tdec-Mmin}) we obtain  
\be
	T_R \approx 6 \times 10^3~\mathrm{GeV} \bigg( \frac{10^{26}~\mathrm{g}}{M_\mathrm{min}} \bigg)^{1/2} \bigg( \frac{M_P}{H_\mathrm{inf}} \bigg)^2,
\ee
where we take $f \theta \approx H_\mathrm{inf}$, $r \approx 1$ and $g_* \approx 100$.
In the case of $\Gamma_I > m_\sigma$, on the other hand, we get the similar relation by simply replacing $T_R$ with $T_\mathrm{osc}$.    
The reheating temperature or the curvaton oscillation temperature is constrained from the inequality (\ref{constraint_on_H_inf2}).
\end{itemize}

\begin{figure}[t]
\centering
\includegraphics [width = 10cm, clip]{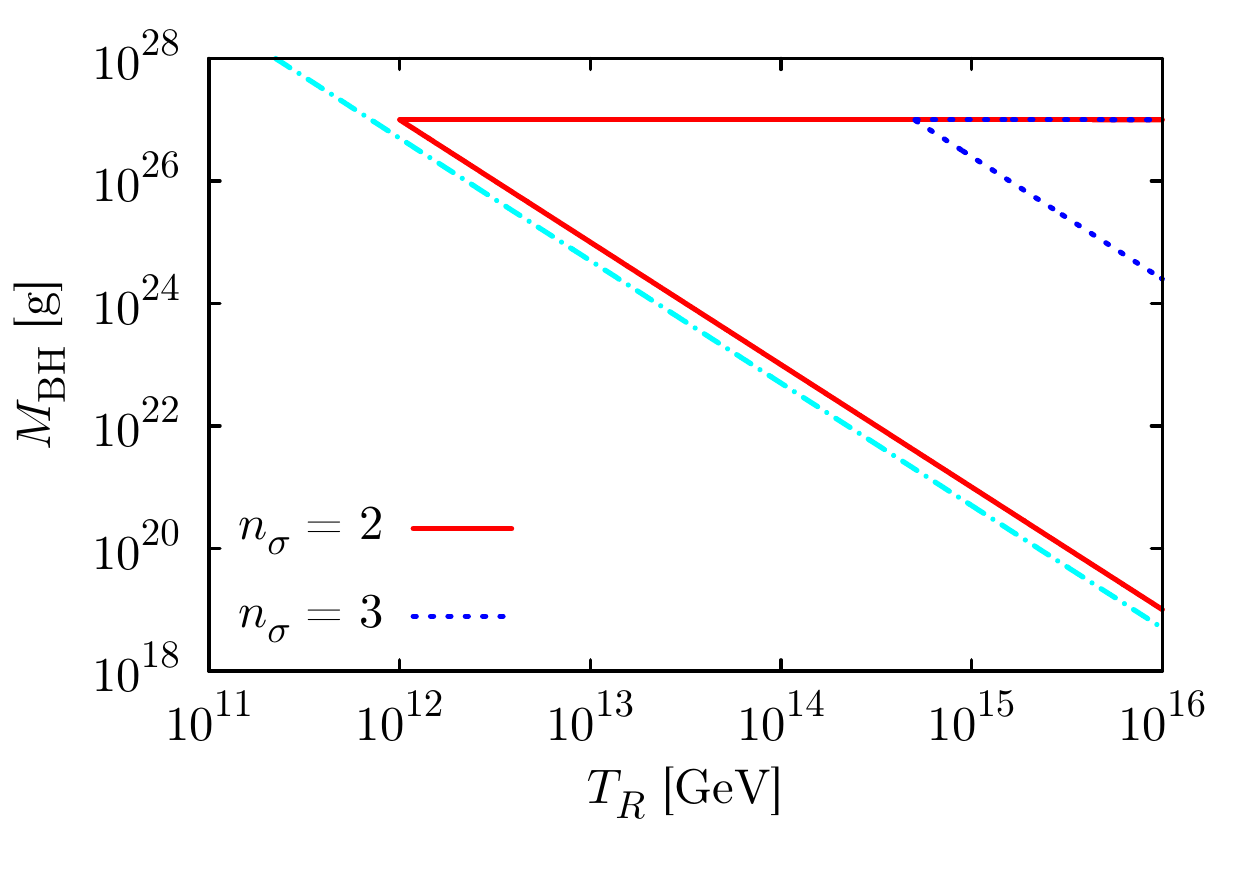}
\caption{
	The allowed parameter region in which PBHs become the dominant component of the CDM in our model is shown.
	The allowed region is inside the respective contours.
	The dotted-blue line and the solid-red line correspond to the boundary in the case of $n_\sigma = $ 3 and 2 respectively.
	The dashed-and-dotted-cyan line is the lower limit on the PBH mass coming from the upper limit on the tensor-to-scalar ratio.
	We have taken $r=1$ and $\theta = 1$.
}
\label{Fig4}
\end{figure}

We show the parameter space allowing for the PBH to take a role of the dominant component of the CDM in Fig.~\ref{Fig4} in the case of $\Gamma_I < m_\sigma$ and $k_\mathrm{c}=1~\mathrm{Mpc^{-1}}$.
The allowed region is inside the respective contours.
The dashed-and-dotted-cyan line is the lower limit on the PBH mass, which comes from the current upper limit on the tensor-to-scalar ratio.
For the PBH to take a role of the dominant component of the CDM, we need the somewhat high reheating temperature $T_R \gtrsim 10^{12}~\mathrm{GeV}$.

Then, we investigate the allowed region of our model parameters, $f$ and $m_\sigma$, in which the current dark matter density can be explained by the PBH.
The allowed parameter region becomes much narrower than Fig.~\ref{Fig4} if we  take into account the decay rate formula (\ref{decay_rate_curv}).
The relation between the decay temperature of the curvaton and PBH mass given by (\ref{Tdec-Mmin}) is rewritten as
\be
	f \simeq 1 \times 10^{14}~\mathrm{GeV}
	~\kappa \bigg( \frac{M_\mathrm{min}}{10^{26}~\mathrm{g}} \bigg)^{1/2} 
	\bigg( \frac{m_\sigma}{10^6~\mathrm{GeV}} \bigg)^{3/2},
\ee
which constrains the allowed region in $f$~--~$m_\sigma$ plane when $10^{17}~\mathrm{g} < M_\mathrm{min} < 10^{27}~\mathrm{g}$ is imposed.
In the case of $m_\sigma > \Gamma_I$, (\ref{reheat1}) is translated into the following inequality : 
\be
	f \gtrsim 1 \times 10^{14}~\mathrm{GeV}~\kappa^{1/3}\bigg( \frac{m_\sigma}{10^{6}~\mathrm{GeV}} \bigg)^{1/3},
	\label{f-m_sigma}
\ee
where we set $\theta = 1$ and $g_* \approx 100$.
In the case of $m_\sigma < \Gamma_I$, on the other hand, the relation from (\ref{reheat2}) becomes approximately same as (\ref{f-m_sigma}) but replacing $\gtrsim$ with $\simeq$.
Moreover, the constraints (\ref{constraint_on_H_inf2}) is trivially translated into upper bound on $f$ by simply replacing $H_\mathrm{inf}$ with $f \theta$.
The constraint (\ref{M_BH_cons}) is rewritten as 
\be
	f \lesssim 1 \times 10^{18} ~\mathrm{GeV}~\kappa \bigg( \frac{M_\mathrm{c}}{10^{34}~\mathrm{g}} \bigg)^{1/2} \bigg( \frac{m_\sigma}{10^6~\mathrm{GeV}} \bigg)^{3/2},
	\label{f-m_sigma2}
\ee
where $M_\mathrm{c}$ is defined as the right-hand-side of (\ref{M_BH_cons}).

\begin{figure}[t]
\centering
\subfigure[]{
	\includegraphics [width = 7.5cm, clip]{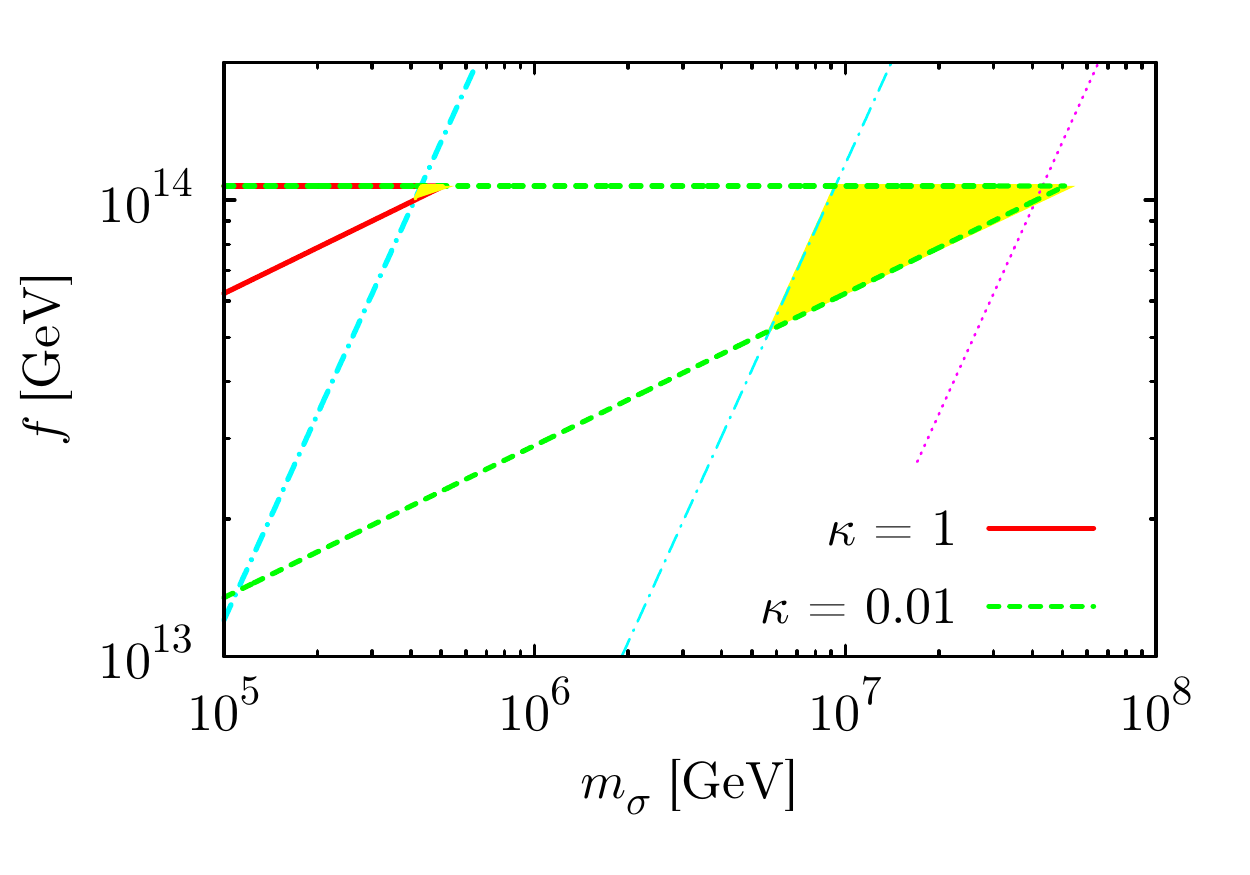}
	\label{Fig5a}
}
\subfigure[]{
	\includegraphics [width = 7.5cm, clip]{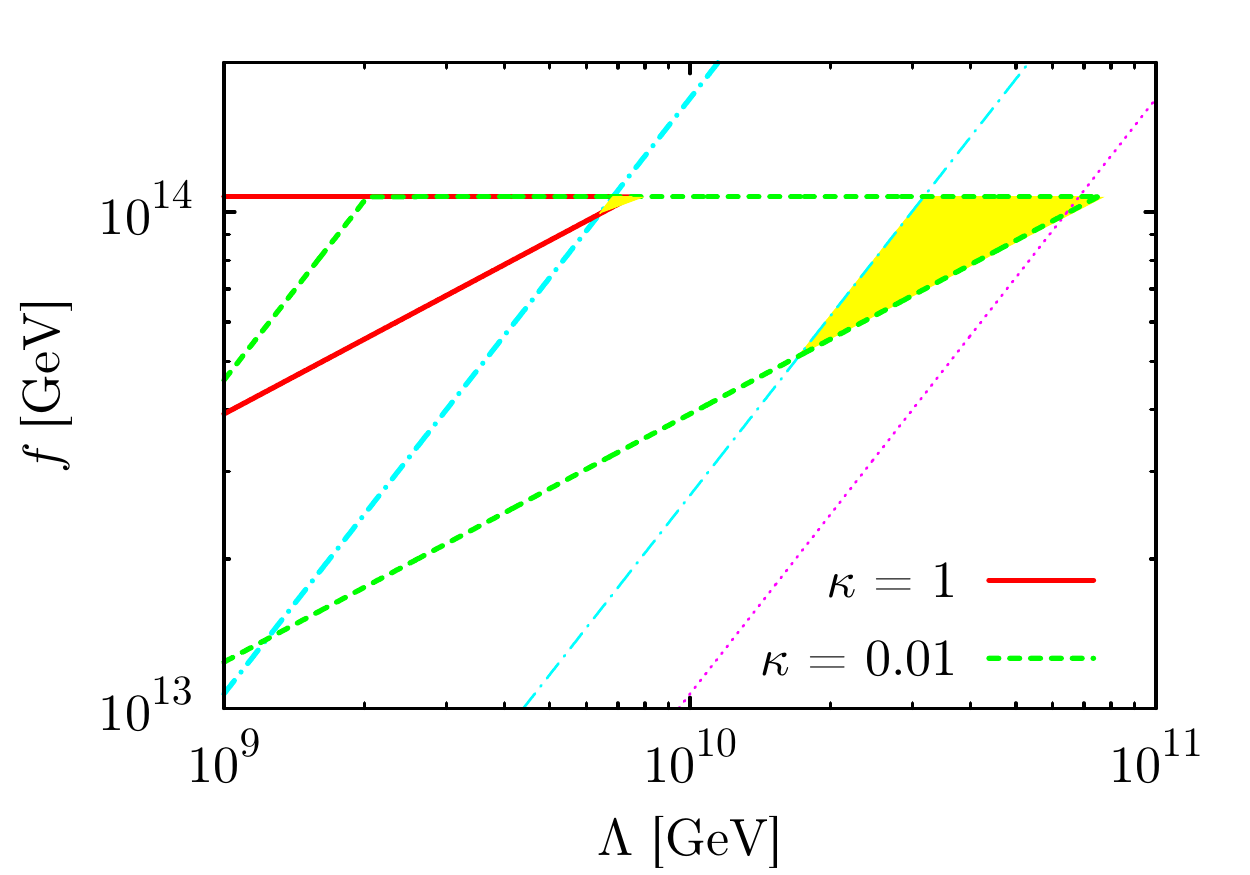}
	\label{Fig5b}
}
\caption{
	The allowed regions for the PBH to be the dominant dark matter in $f$~--~$m_\sigma$ plane and $f$~--~$\Lambda$ plane are shown.
	Inside the thick solid-red (dashed-green) lines, the conditions (\ref{f-m_sigma}) and (\ref{f-m_sigma2}) are satisfied for $\kappa = 1~(0.01)$.
	The thick (thin) dashed-and-dotted-cyan lines corresponds to the upper limit which comes from the maximum mass of PBH dark matter: $M_{\rm BH} = 10^{27}~{\rm g}$ for $\kappa = 1~(0.01)$, 
	so the allowed parameters are inside the yellow shaded regions.
	The thin small-dotted magenta lines correspond to $M_{\rm BH} = 10^{25}~{\rm g}$ for $\kappa = 0.01$.
	We have taken $n_\sigma = 2$ and $\theta = 1$ and assumed $m_\sigma > \Gamma_I$ in both figures.
}
\label{Fig5}
\end{figure}

We summarize the above constraints in Fig.~\ref{Fig5a} and Fig.~\ref{Fig5b} for $n_\sigma = 2$.
In the case of $m_\sigma > \Gamma_I$, the conditions (\ref{f-m_sigma}) and (\ref{f-m_sigma2}) correspond to the region inside the thick solid-red (dashed-green) lines for $\kappa = 1~(0.01)$.
In addition, it must be below the thick (thin) dashed-and-dotted-cyan lines corresponding to the upper bound of the PBH mass, $10^{27}~\mathrm{g}$ for $\kappa = 1~(0.01)$.
Thus, the allowed parameters are inside the yellow shaded regions.
In the opposite case, $m_\sigma < \Gamma_I$, allowed parameters are on the lower  boundary of these regions.
From these, it is found that $f$ and $m_\sigma$ must be $f \sim 5 \times 10^{13}$~--~$10^{14}~\mathrm{GeV}$, $m_\sigma \sim 5 \times 10^{5}$~--~$10^{8}~\mathrm{GeV}$ and $\Lambda \sim 10^{10}$~--~$10^{11}~\mathrm{GeV}$ to explain the current CDM abundance.\footnote{
	Similar results were found when we considered the PBH formation in the matter (curvaton) dominated era.
	For example, setting $r = 10$ and same parameters as those we have taken in Fig.~\ref{Fig5}, 
	we found $f \sim 3 \times 10^{14}~{\rm GeV}$ and $m_\sigma \sim 3 \times 10^{5} - 10^6~{\rm GeV}$ for $n=2$ to explain the CDM abundance.
}

\begin{figure}[t]
\centering
\subfigure[]{
	\includegraphics [width = 7.5cm, clip]{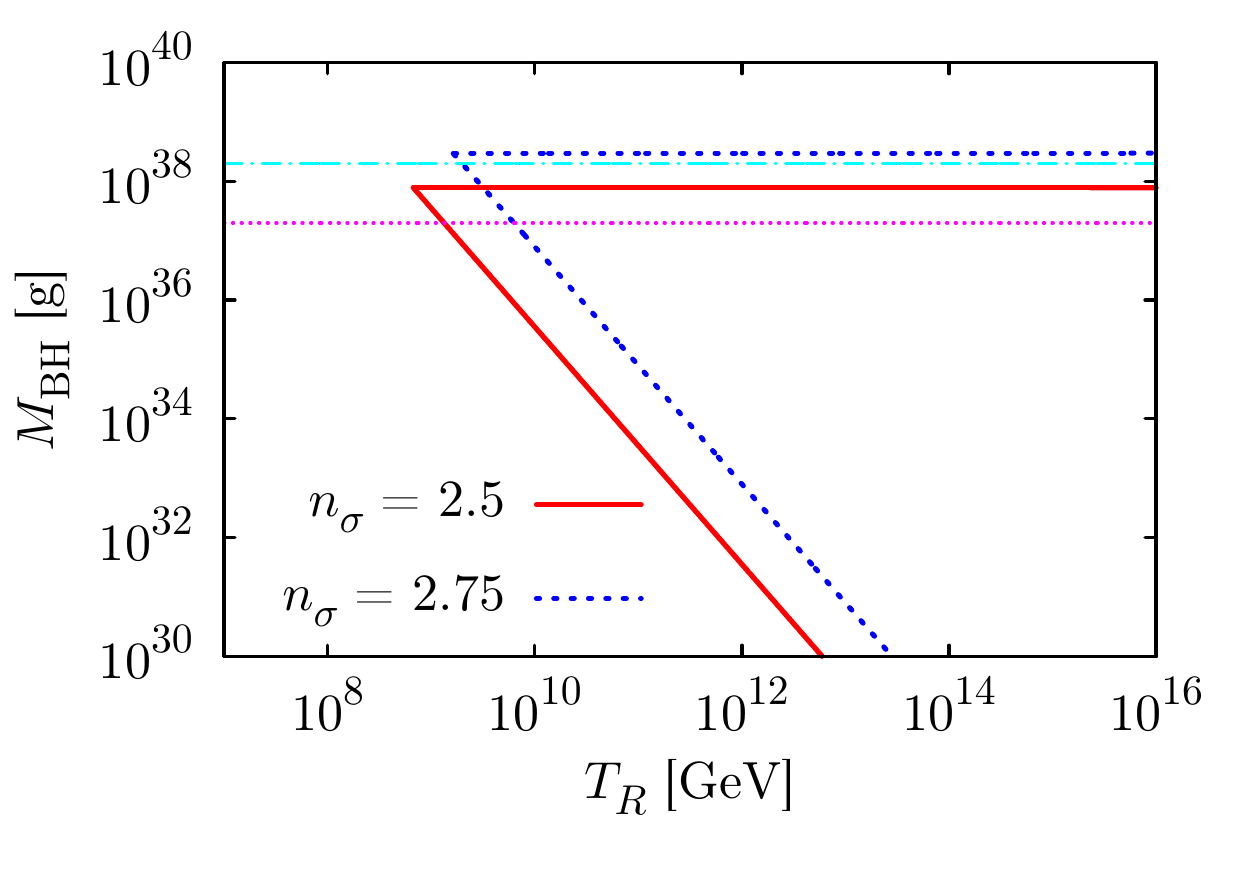}
	\label{Fig6a}
}
\subfigure[]{
	\includegraphics [width = 7.5cm, clip]{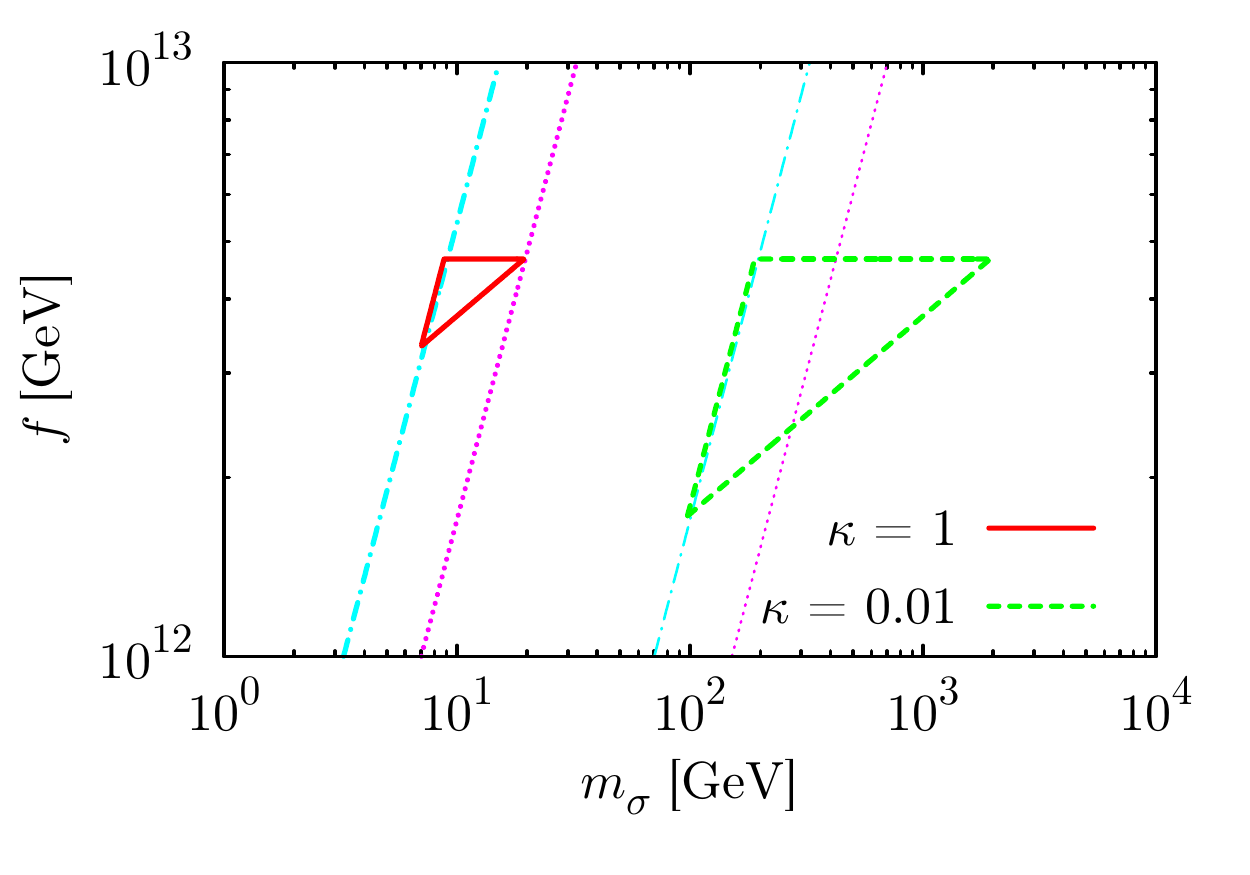}
	\label{Fig6b}
}
\caption{
	The allowed regions for the PBH to be the seed of the SMBHs in $M_\mathrm{BH}$~--~$T_R$ plane and $f$~--~$m_\sigma$ plane are shown in Fig.~\ref{Fig6a} and Fig.~\ref{Fig6b} respectively.
	In Fig.~\ref{Fig6a}, the allowed region is inside the solid-red line (the dotted-blue line) for $n_\sigma = 2.5$ ($n_\sigma = 2.75$) 
	and the dashed-and-dotted-cyan line (small-dotted-magenta line) corresponds to $M_\mathrm{BH} = 10^5M_\odot$ ($10^4 M_\odot$).
	In Fig.~\ref{Fig6b}, the allowed region is inside the solid-red line (the dashed-green) 
	and the thick (thin) dashed-and-dotted-cyan line and small-dotted-magenta line correspond to $M_\mathrm{BH} = 10^5 M_\odot$ and $M_\mathrm{BH} = 10^4 M_\odot$ respectively for $\kappa = 1$ ($0.01$).
	We have taken $r=1$ and $\theta = 1$ in both figures and $n_\sigma = 2.75$ in Fig.~\ref{Fig6b}. 
}
\label{Fig6}
\end{figure}

Another outcome of our model is the possibility of explaining the seeds of SMBHs.
The initial mass fraction of PBHs as seeds of SMBHs is constrained by the observed comoving number density of QSOs: $a^3 n_\mathrm{QSO} \simeq (6 \pm 2) \times 10^{-10}~\mathrm{Mpc}^{-3}$ \cite{Willott:2003xf}.
The comoving number density of PBHs is given by
\be
	a^3 n_\mathrm{PBH} \simeq 6 \times 10^{18} \beta ~\mathrm{Mpc}^{-3}~\bigg( \frac{g_*}{10} \bigg)^{-1/4} \bigg( \frac{M_\odot}{M_\mathrm{BH}} \bigg)^{3/2},
\ee
so, compared with $a^3 n_\mathrm{QSO}$, $\beta$ is estimated as
\be
	\beta \sim 2 \times 10^{-21} \bigg( \frac{g_*}{10} \bigg)^{1/4} \bigg( \frac{M_\mathrm{BH}}{10^{5}~M_\odot} \bigg)^{3/2}.
\ee
Since a quite large mass and a narrow mass spectrum of the PBH are needed to explain the SMBH, we set $M_\mathrm{min} \sim M_*$ which leads $\alpha \simeq 0.1$, 
$H_\mathrm{inf} / f \theta \sim 2$ and $\mathcal P_{\zeta,\mathrm{curv}}(k_*) \sim 1 \times 10^{-2}$.
Then the parameter space is constrained by the same way as those of the PBH dark matter case and we summarize it in Fig.~\ref{Fig6}.

In Fig.~\ref{Fig6a}, the allowed region is inside the solid-red line (the dotted-blue line) for $n_\sigma = 2.5$ ($n_\sigma = 2.75$).
The dashed-and-dotted-cyan line (the small-dotted-magenta line) corresponds to $M_\mathrm{BH} = 10^5 M_\odot$ ($10^4 M_\odot$), on which the SMBH is explained by the PBH.
In Fig.~\ref{Fig6b}, the allowed region is inside the solid-red line (the dashed-green line) for $\kappa = 1$ ($0.01$).
The thick (thin) dashed-and-dotted-cyan line and small-dotted-magenta line correspond to $M_\mathrm{BH} = 10^5 M_\odot$ and $M_{\rm BH} = 10^4 M_\odot$ respectively for $\kappa = 1$ ($0.01$).
We found that our model can provide the seeds of SMBHs for $T_R \gtrsim 10^9~\mathrm{GeV}$, $f \sim 10^{12}~\mathrm{GeV}$, $m_\sigma \sim 0.5$~--~$100~\mathrm{GeV}$ and $\Lambda \sim 10^6$~--~$10^7$~GeV.

However, the SMBHs cannot be a significant part of the dark matter density of the universe.
Fortunately, various axion-like particles often appears in particle physics theories.
One of them may be the curvaton which is responsible for SMBHs as discussed above. 
Another axion field can play a role of the usual QCD axion which solves the strong CP problem.
If the Peccei-Quinn scale $f_a$ is  $\sim 10^{12}$~GeV, the QCD axion can account for the dark matter of the universe.
The coincidence of two independent scales $f\simeq f_a \sim 10^{12}$~GeV may be very interesting.
Furthermore, it is pointed out that the axion dark matter is a good candidate consistent with the presence of the primordial SMBHs~\cite{Bringmann:2011ut}. 
The required scale $\Lambda \sim 10^6$~--~$10^7$~GeV is coincide with the SUSY breaking scale when it is mediated by gauge interactions, which suggests that the dynamics generates the curvaton mass  may be related to physics of SUSY breaking.


\section{Conclusion} \label{conc}

We considered the axion-like curvaton model based on the SUSY, in which the curvaton is identified as the phase direction contained in some complex scalar field. 
Because of the Hubble-induced mass of the radial part of the complex scalar field, the power spectrum of the curvature perturbations from the curvaton becomes extremely blue such as $n_\sigma = 2$~--~$4$.
In order not to contradict with the WMAP observation of the spectral index, in our model, the large scale perturbations ($k \lesssim 1~\mathrm{Mpc}^{-1}$) are generated by the inflaton 
giving the almost scale-invariant power spectrum and the contribution from the curvaton becomes significant at sufficiently  small scales.
We showed that, by use of such a extremely blue spectrum, the PBHs are formed from the collapse of the overdensity regions and the produced PBHs have a peaked mass spectrum.
It is found that, in a certain parameter region, PBHs with mass $10^{17}-10^{27}$~g can eventually become the dominant component of the CDM.
Furthermore, it is found that the PBHs with quite large masses ($\sim 10^{5}M_{\odot}$)   and very narrow mass spectrum can be formed and these can be the seeds of SMBHs.

In this paper we have derived the scalar potential in the frame work of supergravity. 
However, we can build the model without supersymmetry if we start with the potential~(\ref{scalar_pot}). 
The Hubble induced mass terms~(\ref{Hubble_mass}) which are necessary for generating the blue-tilted power spectrum can be obtained through couplings with the inflaton field.   
For example, suppose that a scalar $\varphi$ causes chaotic inflation and its potential is given by $V(\varphi) =\lambda \varphi^4$. 
Then the term like $g\varphi^2 |\Phi|^2$ ($g$: small coupling) lead to the Hubble induced mass term for $\Phi$ if we take appropriate $g$.

\section*{Acknowledgment}

We thank Fuminobu Takahashi for useful discussions. 
This work is supported by Grant-in-Aid for Scientific research from
the Ministry of Education, Science, Sports, and Culture (MEXT), Japan,
No.\ 14102004 (M.K.), No.\ 21111006 (M.K.) and also 
by World Premier International Research Center
Initiative (WPI Initiative), MEXT, Japan. 
N.K. is supported by the Japan Society for the Promotion of Science (JSPS).

  

\end{document}